\documentclass[structabstract]{aa}
\usepackage{graphicx}
\usepackage{txfonts}
\usepackage{supertabular}
\usepackage[modulo,switch]{lineno} 
\usepackage{natbib}
\bibpunct{(}{)}{;}{a}{}{,} 

\begin{document}
\def\cd  {{$\mbox{cd}^{-1}$}}
   \title{Pulsations in the late-type Be star HD 50\,209 detected by CoRoT
   \thanks{Based on observations made with the CoRoT satellite, with FEROS at the 2.2m telescope of the La Silla Observatory under the ESO Large Programme LP178.D-0361 and with Narval at the T\'{e}lescope Bernard Lyot of the Pic du Midi Observatory.}\fnmsep 
   \thanks{Table~\ref{tablafreqscurta} is only available in electronic form via http://www.edpsciencies.org}}


   \author{P.~D. Diago\inst{1,}\thanks{Current address: Valencian International University (VIU), Jos\'{e} Pradas Gallen s/n, 12006 Castell\'{o}n, Spain.}
        \and
        J. Guti\'{e}rrez-Soto\inst{2,3}
	\and
	M. Auvergne\inst{3}
	\and 
	J. Fabregat\inst{1}
	\and
	A.-M. Hubert\inst{2}
	\and
	M. Floquet\inst{2}
	\and
	Y. Fr\'{e}mat\inst{2,4}
	\and
	R. Garrido\inst{5}
	\and
	L. {Andrade}\inst{6}
	\and
	B. de Batz\inst{2}
	\and
	M. Emilio\inst{7}
	\and
	F. Espinosa-Lara\inst{2}
	\and
	A.-L. Huat\inst{2}
	\and
	E. Janot-Pacheco\inst{6}
	\and
	B. Leroy\inst{3}
	\and	
	C. Martayan\inst{2,4}
	\and
	C. Neiner\inst{2}
	\and
	T. Semaan\inst{2}
	\and
	J. Suso\inst{1}
	\and
	C. Catala\inst{3}
	\and
	E. Poretti\inst{8}
	\and
	M. Rainer\inst{8}
	\and
	K. Uytterhoeven\inst{8,9,}\thanks{Current address: Laboratoire AIM, CEA/DSM-CNRS-Universit\'e Paris Diderot; CEA, IRFU, SAp, centre de Saclay, F-91191, Gif-sur-Yvette, France.}
	\and
	E. Michel\inst{3}
	\and
	R. Samadi\inst{3}.}

   \institute{
   Observatori Astron\`{o}mic de la Universitat de Val\`{e}ncia, Edifici Instituts d'Investigaci\'{o},  Pol\'{i}gon La Coma, 46980 Paterna, Val\`{e}ncia, Spain. \email{Pascual.Diago@uv.es}
    \and 
	     GEPI, Observatoire de Paris, CNRS, Universit\'e Paris Diderot, Place Jules Janssen 92195 Meudon Cedex, France.
	\and 
	     LESIA, Observatoire de Paris, CNRS, Universit\'e Paris Diderot, Place Jules Janssen 92195 Meudon Cedex, France.
	\and 
	     Royal Observatory of Belgium, 3 Avenue Circulaire, B-1180 Brussels, Belgium.
	\and 
	     Instituto de Astrof\'{i}sica de Andaluc\'{i}a (CSIC), Camino Bajo de Hu\'{e}tor, 24, 18008, Granada, Spain.
	\and 
	     Universidade de S\~{a}o Paulo, Instituto de Astronomia, Geof\'{i}sica e Ci\^{e}ncias Atmosf\'{e}ricas - IAG, Departamento de Astronomia, Rua do Mat\~{a}o, 1226 - 05508-900. S\~{a}o Paulo, Brazil.
	\and 
	     Observat\'{o}rio Astron\^{o}mico/DEGEO, Universidade Estadual de Ponta Grossa, Av. Carlos Cavalcanti, 4748 Ponta Grossa, Paran\'{a}.
	\and 
	     INAF-Osservatorio Astronomico di Brera, Via E. Bianchi 46, 23807 Merate (LC), Italy.
	\and 
	     Instituto de Astrof\'{\i}sica de Canarias, Calle Via L\'{a}ctea s/n, E-38205 La Laguna (TF), Spain.
             }

   \date{Received / accepted }

 
  \abstract
   {The presence of pulsations in late-type Be stars is still a matter of controversy. It constitutes an important issue to establish the relationship between non-radial pulsations and the mass-loss mechanism in Be stars.}
   {To contribute to this discussion, we analyse the photometric time series of the B8IVe star HD 50\,209 observed by the CoRoT mission in the seismology field.}
   {We use standard Fourier techniques and linear and non-linear least squares fitting methods to analyse the CoRoT light curve. In addition, we applied detailed modelling of high-resolution spectra to obtain the fundamental physical parameters of the star.}
   {We have found four frequencies which correspond to gravity modes with azimuthal order $m=0,-1,-2,-3$ with the same pulsational frequency in the co-rotating frame. We also found a rotational period with a frequency of $0.679$ \cd (7.754 $\mu$Hz).}
   {HD 50\,209 is a pulsating Be star as expected from its position in the HR diagram, close to the SPB instability strip.}

%
%
\maketitle
%

\section{Introduction}

Classical Be stars are defined as main sequence or slightly evolved B-type stars whose spectrum has or had at some time one or more Balmer lines in emission. They are physically understood as rapidly rotating B-type stars with line emission arising from a circumstellar disk in the equatorial plane, composed of matter ejected from the stellar photosphere by mechanisms not yet understood \citep[see][for a complete review]{2003PASP..115.1153P}. A significant fraction of Be stars show short-term photometric and spectroscopic periodic variability, which is commonly attributed to non-radial pulsations. As Be stars occupy the same region of the HR diagram as $\beta$ Cephei and SPB stars, it is generally assumed that pulsations have the same origin, i.e. p- and/or g-mode pulsations driven by the $\kappa$-mechanism associated with the Fe bump. The current theoretical models have difficulties in describing the pulsational characteristics of Be stars, due to the high rotational velocity of these objects. A brief description of the current knowledge of the pulsational behavior of Be stars can be found in \citet{Neiner2009} and \citet{Emilio2009}.


HD 50\,209 is a late Be star of spectral type B8IVe and magnitude $V = 8.36$. It has been studied by \citet{2007A&A...476..927G}, using data from Hipparcos, ASAS-3 and the OSN (Observatorio de Sierra Nevada). The Hipparcos data analysis yielded a frequency of variation at 1.689 \cd considered as uncertain and another frequency at 1.47 \cd, although with a lower amplitude. From OSN, the analysis revealed a frequency at 1.4889 \cd. The analysis of the ASAS-3 dataset showed significant peaks at frequency 2.4803 \cd and 1.4747 \cd.

The occurrence of pulsations in late B-type stars has been a matter of controversy in the recent literature. \citet{1998A&A...335..565H} showed that pulsations in B6-B9 type stars are much less common than in their early-type counterparts. \citet{1989A&A...222..200B} failed to detect line profile variations in the spectra of B8-B9.5 stars. However, \citet{2007ApJ...654..544S} presented the detection of low amplitude g-modes in the B8Ve star $\beta$ CMi. To ascertain whether Be stars of all types do present pulsations is a key issue in order to establish the relation between non-radial pulsations and the mass ejection mechanisms.

The CoRoT satellite \citep[][]{2009arXiv0901.2206A} observed HD 50\,209 in its seismology field. The observations span 136 days in the Galactic anti-centre direction (LRA1), between October 18th 2007 and March 3rd 2008, with a sampling of 32 seconds. The light curve contains 328\,279 data-points with a duty cycle of $89\%$.

\section{Frequency analysis}


For the frequency analysis, we employed the code \textit{pasper} \citep[][]{2008A&A...480..179D}, which is based on the classical discrete Fourier transform \citep{1975Ap&SS..36..137D, 1982ApJ...263..835S} and the least-square fitting of a sinusoidal function in the time domain. When a frequency is found, it is prewhitened from the original data and a new step starts looking for a new frequency in the residuals. The method is iterative and stops when the new frequency is not statistically significant.

The criterion used to determine whether the frequencies are statistically significant is the \emph{signal to noise amplitude ratio requirement} described in \citet{1993A&A...271..482B}. It consists of the calculation of the SNR of the frequency in the periodogram. This is made calculating the signal as the amplitude of the peak for the frequency obtained and the noise as the average amplitude in the residual periodogram after the prewhitening of all the frequencies detected. \citet{1993A&A...271..482B} proposed that a value of SNR $\geq4$ is a reliable criterion to distinguish between peaks due to real frequencies and noise.

All the techniques used in this paper are described in detail by \citet{G-S2009}. As mentioned in that paper, the orbital characteristics of the CoRoT satellite produces signal in the data at the orbital frequency ($13.97$ \cd, 161.689 $\mu$Hz) and day/night peaks ($2.007$ \cd, 23.229 $\mu$Hz) and their harmonics. All those frequencies of instrumental origin have been removed from the list of detected frequencies.


The resolution in frequency of our analysis is $1/2\,T \sim 3.66 \times 10^{-3}$ \cd \citep[][]{2008A&A...481..571K}, $T$ being the total interval covered by observations. The uncertainty on the detected frequencies is $\sqrt{3} \times 10^{-6}$ \cd. This value has been derived analytically using the formula given by \citet{1999DSN-M&O}, and taking into account the correlations in the residuals, as described by \citet{1991MNRAS.253..198S}.

\begin{table}
\caption{Principal frequencies of each set detected in HD 50\,209.}
\label{tablasets}
\centering
\begin{tabular}{c c c c c c}
\hline
\hline
Set	& Freq.	  & Freq.	& Amp.	& Main freq.& Comment.	\\
	& [\cd]	  & [$\mu$Hz]	& [mmag]& 	    &		\\
\hline
$F_{1}$ & 1.48444 & 17.181	& 2.529   & $f_{1}$ &     $f_{5}+2\, f_{4}$\\
 	& 1.49028 & 17.248	& 1.616   &       &	       \\
 	& 1.47860 & 17.113	& 1.090   &       &	       \\
 	& 1.49613 & 17.316	& 0.653   &       &	       \\
\hline
$F_{2}$	& 2.16238 & 25.027	& 0.904   & $f_{2}$ &     $f_{5}+3\, f_{4}$\\
 	& 2.16822 & 25.095	& 0.495   &       &	       \\
\hline
$F_{3}$	& 0.79482 & 9.199	& 0.607    & $f_{3}$ &     $f_{5}+f_{4}$\\
 	& 0.77874 & 9.013	& 0.407    &       &	       \\
 	& 0.80650 & 9.334	& 0.358    &       &	       \\
\hline
$F_{4}$	& 0.67939 & 7.863	& 0.425    & $f_{4}$ & $f_{rot}$\\
 	& 0.69108 & 7.998	& 0.368    &       &	       \\
\hline
$F_{5}$	& 0.10811 & 1.251	& 0.202    & $f_{5}$ &	       \\
\hline
$F_{6}$	& 2.96889 & 34.362	&  0.121  & $f_{6}$ &  $2\, f_{1}$\\
\hline
\end{tabular}
\end{table}

\begin{figure}
	\resizebox{\hsize}{!}{\includegraphics{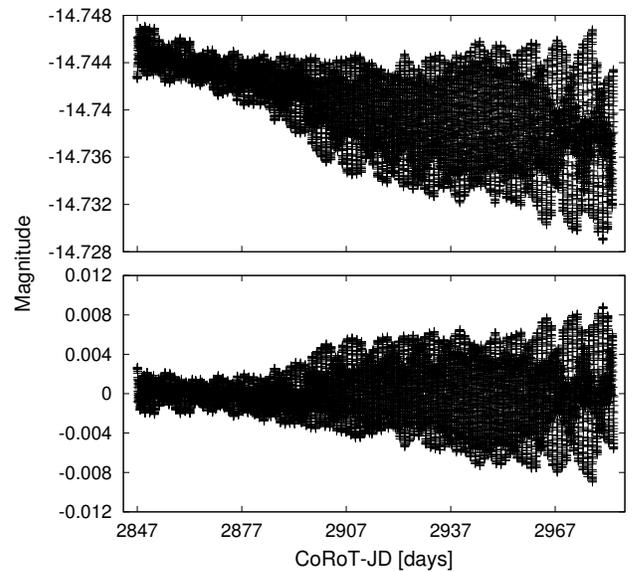}}
   	\caption{\textit{Top:} CoRoT light curve of the star HD 50\,209. \textit{Bottom:} CoRoT light curve  detrended with a polynomial of second degree.}
   	\label{LC}
\end{figure}

\begin{figure*}
	\resizebox{\hsize}{!}{\includegraphics{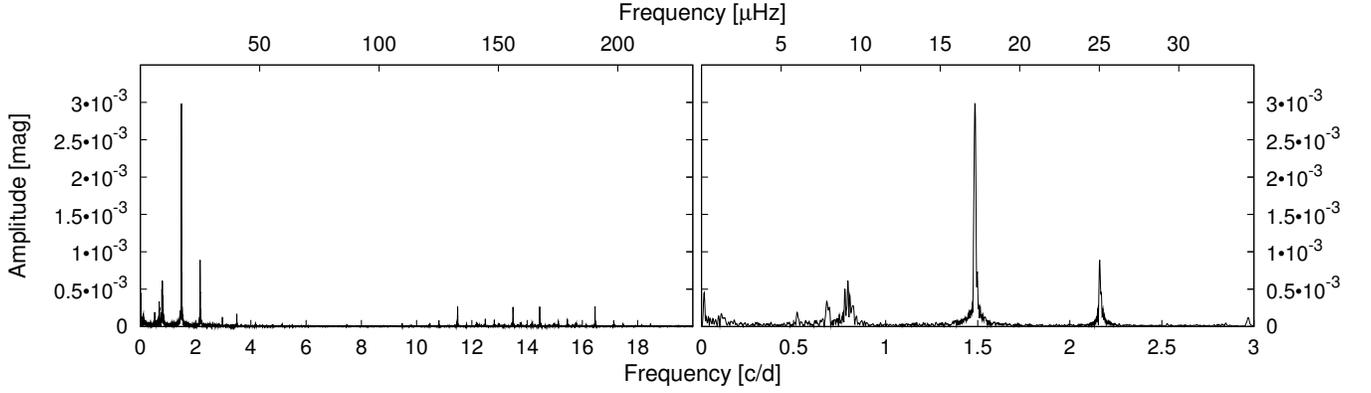}}
   	\caption{\textit{Left:} Complete Fourier spectrum for HD 50\,209 showing aliasing effects at the orbital frequency 13.97 \cd. \textit{Right:} Fourier spectrum for HD 50\,209 in the domain of the detected stellar frequencies.}
   	\label{DFT}
\end{figure*}

\begin{figure}
	\resizebox{\hsize}{!}{\includegraphics{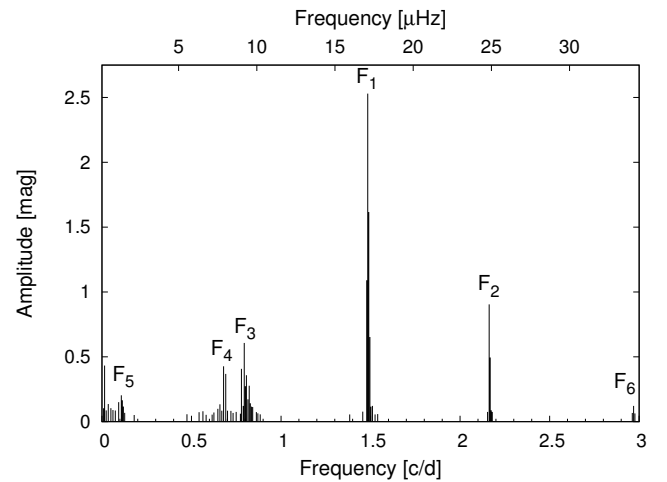}}
   	\caption{The 60 significant frequencies detected for the star HD 50\,209 revealing six different and separated frequency groups.}
   	\label{peaks}
\end{figure}

In Fig.~\ref{LC} (top panel) we present the CoRoT light curve of HD 50\,209. A long-term decreasing pattern is apparent. It is a common feature of the CoRoT light curves to show a linear or an almost linear decreasing pattern of instrumental origin due to the CCD ageing. On the other hand, most Be stars present long-term variations which in our case contribute to the observed variability. We have removed these trends by fitting a polynomial function of second degree without making assumptions on their origin, obtaining the light curve depicted in the bottom panel of Fig.~\ref{LC}.


\begin{table*}
\caption{Complete list of significant frequencies obtained with \textit{pasper} for the star HD 50\,209. The time reference for the phase is HJD = 2454391.95. The full table is available online.}
\label{tablafreqscurta}
\centering
\begin{tabular}{c c c c c c c c c c}     
\hline\hline
\#	 & Freq.  		& Freq.	 	& Amp.		  & Amp. error	& Phase	 		& Phase error     & S/N   		   & Comment.\\
	 & [\cd]  		& [$\mu$Hz]	& [mmag]	  & [mmag]		& [0,1]	 		& [0,1]	 		  &	 			   &	   \\
\hline
1    &    1.48444   &  17.181   &   2.5299    &   0.0011    &   0.7348	    &	 0.0001   &    265      &  $f_{1}$     \\
2    &	  1.49028   &  17.2486  &   1.6167    &   0.0012    &   0.1577	    &	 0.0001   &    169      &       \\
3    &	  1.47860   &  17.1134  &   1.0901    &   0.0011    &   0.2747	    &	 0.0002   &    114      &       \\
4    &	  2.16238   &  25.0275  &   0.9043    &   0.0010    &   0.9746	    &	 0.0002   &    94       &  $f_{2}$     \\
5    &	  0.79482   &  9.19931  &   0.6077    &   0.0011    &   0.9866	    &	 0.0003   &    63       &  $f_{3}$     \\
$\cdots$& $\cdots$	&$\cdots$	&$\cdots$	  &$\cdots$		&$\cdots$		&$\cdots$	  &$\cdots$		&$\cdots$		\\	
\hline
\end{tabular}
\end{table*}

In Fig.~\ref{DFT} we display the Fourier spectrum, showing the typical aliasing at the orbital frequency \citep[as described in][]{G-S2009}. From the frequency analysis, we obtain 60 significant frequencies which are listed in Table~\ref{tablafreqscurta} (full table only available online). Note that frequencies around $3/T\sim0.022$ \cd (0.255 $\mu$Hz) or lower correspond to periods of the order of the entire dataset time-span, and hence should be considered with caution. They might be produced by the detrending process or artifacts of the frequency analysis. However, we have included them for completeness.

\begin{figure}
	\resizebox{\hsize}{!}{\includegraphics{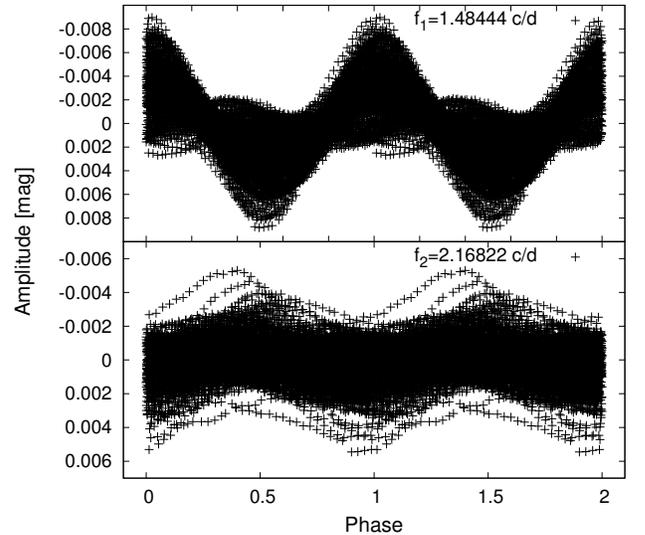}}
   	\caption{\textit{Top:} Phase diagram for the HD 50\,209 data folded with the frequency $f_{1}=1.48444$ \cd (17.181 $\mu$Hz). \textit{Bottom:} Phase diagram for the HD 50\,209 data, prewhitened from the frequency $f_{1}$ and folded with the frequency $f_{2}=2.16238$ \cd (25.027 $\mu$Hz). Note that two cycles are shown for clarity.}
   	\label{phases}
\end{figure}

The entire set of detected frequencies together with their amplitudes is plotted in Fig.~\ref{peaks}. The frequencies found are distributed in 6 main groups ($F_{i}$ with $i=1,...,6$). The frequencies with the highest amplitudes are listed in Table~\ref{tablasets}. All groups are clearly separated, except those centered on $f_{3}$ and $f_{4}$. In each group we note the existence of equidistant multiplets separated by intervals marginally larger than our frequency resolution. In Fig.~\ref{phases} we provide the diagrams of the folded light curve with the two main frequencies $f_{1}$ and $f_{2}$.

\begin{figure}
	\resizebox{\hsize}{!}{\includegraphics{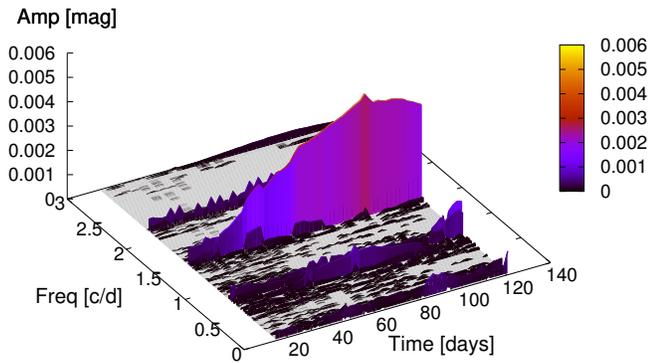}}
   	\caption{3-D view of the frequencies determined using the time-frequency analysis with a sliding window of 30 days. The color scale indicates the amplitude of the detected frequencies in HD 50\,209. The time origin corresponds to  HJD = 2454391.95.}
   	\label{3D}
\end{figure}

In addition, we performed a time-frequency analysis of the original light curve by applying the \textit{pasper} code to sliding windows with different sizes \citep[for details on the method, see][]{Huat2009}. The results presented in Fig.~\ref{3D} show that all the main frequencies show large amplitude changes. See Section~\ref{discussion} for a detailed discussion of these changes.

\section{Ground-based observations}\label{ground-based}

High resolution spectroscopy of HD 50\,209 was obtained with the FEROS spectrograph at the 2.2m telescope in La Silla, in the framework of a large program complementary to the CoRoT mission. Observations were carried out one year prior to the LRA1 run of CoRoT to investigate spectroscopically the rapid variability previously detected in photometry by \citet{2007A&A...476..927G}. Seventy spectra with high spectral resolution and high signal-to-noise ratio were obtained. Fourteen additional spectra were obtained with the Narval spectropolarimeter at the TBL at Pic du Midi Observatory (France), contemporary to the CoRoT run. However, due to bad weather conditions, the signal-to-noise ratio is not high enough to search for low amplitude variability in the Narval data.

FEROS observations ($R=48000$, $3600-9200$ \AA{}) were reduced with MIDAS (wavelength calibration, bias and flat field corrections and earth motion correction). For the Narval observations ($R=65000$) the sum of the 4 sub-exposures obtained in Stokes V sequences were used. Exposures were reduced locally with LibrEsprit based on the Esprit software \citep[][]{1997MNRAS.291..658D} and then summed. The continuum normalization was carried out with IRAF.


The fundamental physical parameters of HD 50\,209 have been accurately determined from the newly available spectroscopic data adopting a procedure described by \citet{2006A&A...451.1053F} and correcting for gravitational darkening effects \citep[][]{2005A&A...440..305F}. The obtained parameters are derived by fitting to the spectra, adopting the different values of $\Omega/\Omega_{c}$ presented in the first row of Table~\ref{fp}. The results are presented in Table~\ref{fp} and are consistent with the previous parameter determination \citep[][]{2006A&A...451.1053F}.

\begin{table*}
\caption{Fundamental parameters for HD 50\,209 computed for different values of $\Omega/\Omega_{c}$ and corrected for veiling. The errors in $T_{eff}$, $\log g$ and $V\, \sin\,i$  are taken from the apparent parameters.}
\label{fp}
\centering
\begin{tabular}{c c c c c c c c c}
\hline
\hline
$\Omega/\Omega_{c}$& $i$	& $T_{\rm eff}$	 & $\log g$	 & $V\, \sin\,i$& $M$		& $L$		& $R_{\rm eq}$ 	& $f_{\rm rot}$	\\
		   & [deg] 	& [K]		 & [cgs]	 & [km/s]	& [$M_{\odot}$] &[$L_{\odot}$]	&[$R_{\odot}$]  & [\cd]		\\
\hline
0.80		& $89\pm20$ 	& $13600\pm 1500$& $3.60\pm 0.11$& $192\pm 20$	& $4.67\pm0.95$ & $2.97\pm0.39$ & $6.44\pm1.53$ & $0.60\pm0.20$ \\
0.90		& $64\pm14$ 	& $13600\pm 1500$& $3.56\pm 0.11$& $205\pm 20$	& $4.76\pm1.02$ & $3.02\pm0.39$ & $7.27\pm1.74$ & $0.62\pm0.20$ \\
0.95		& $57\pm15$ 	& $13600\pm 1500$& $3.53\pm 0.11$& $211\pm 20$	& $5.07\pm1.15$ & $3.08\pm0.40$ & $8.06\pm2.03$ & $0.62\pm0.22$ \\
0.99		& $55\pm12$ 	& $13400\pm 1500$& $3.46\pm 0.11$& $221\pm 20$	& $4.99\pm1.22$ & $3.14\pm0.41$ & $9.34\pm2.35$ & $0.58\pm0.20$ \\
\hline
\end{tabular}
\end{table*}

Spectroscopic data also show that emission is present in the first Balmer lines and visible up to H$\delta$. H$\alpha$ is a strong emission line ($EW=-19$ \AA{} , $Imax=4.68$ Icont) with inflection points on the wings and close double peaks at the centre. The line profile is typical of a Be star seen under a moderate inclination angle \citep[][]{1996A&AS..116..309H}. H$\beta$, H$\gamma$ and H$\delta$ show a symmetrical and double-peaked emission embedded in the broad wings of the photospheric profile. Moreover, numerous \ion{Fe}{ii} and \ion{Cr}{ii} lines as well as the IR \ion{Ca}{ii}  triplet and \ion{O}{i} 8446 $\AA$ lines are in emission with a double component. Note that we have not found significant changes in the global emission state of HD 50\,209 between the FEROS and Narval spectra taken about one year apart.

\begin{figure}
	\resizebox{\hsize}{!}{\includegraphics{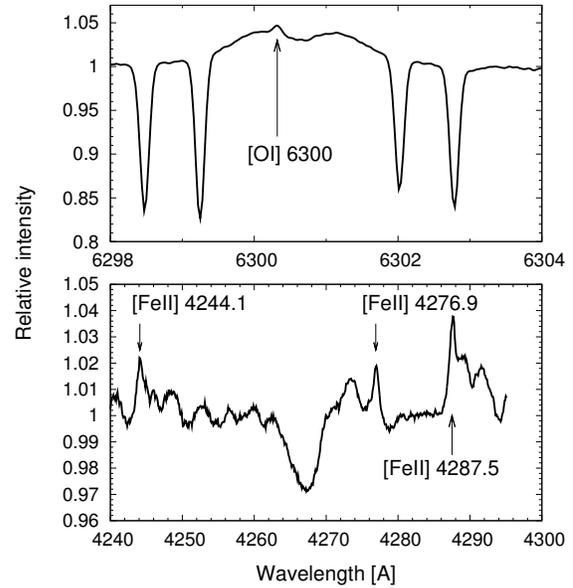}}
   	\caption{\textit{Top:} Forbidden emission line of [\ion{O}{i}] at 6300 $\AA$. \textit{Bottom:} Three forbidden emission lines of [\ion{Fe}{ii}] at 4244.1 $\AA$ (21F), 4276.9 $\AA$ (21F) and 4287.5 $\AA$ (7F) from the spectra of HD 50\,209.}
   	\label{forblines}
\end{figure}

In addition, forbidden single-peaked [\ion{O}{i}] 6300 $\AA$ and [\ion{Fe}{ii}] emission lines are detected with a very weak intensity ($I\leq 1.01 \%$ of the continuum, see Fig.~\ref{forblines}). The star is isolated, outside any formation region, and without the large IR excess typical of HAeBe stars. It could be an extreme Be star according to \citet{1998ASSL..233..269D} or an unclB[e] star in the scheme proposed by \citet{1998A&A...340..117L}, though the B[e] character is very weak.

\begin{figure}
	\resizebox{\hsize}{!}{\includegraphics{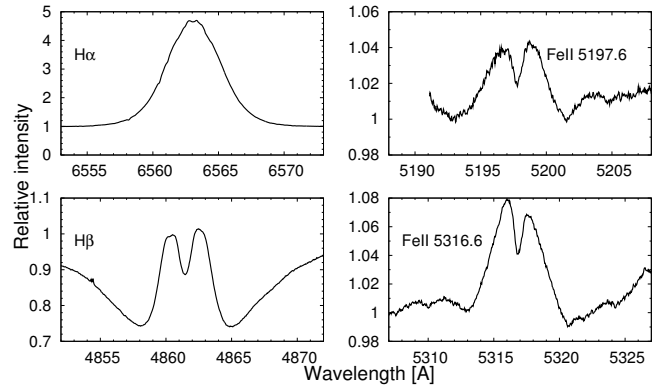}}
   	\caption{\textit{Left panels:} H$\alpha$ and H$\beta$ emission. \textit{Right panels:} Different emission lines in the \ion{Fe}{ii} region in the star HD 50\,209. In all panels the horizontal axis is wavelength and the vertical one is relative intensity.}
   	\label{Halpha}
\end{figure}

The mean H$\alpha$, H$\beta$ and \ion{Fe}{II} 5197.6 $\AA$ and 5316.6 $\AA$ profiles are shown in Fig.~\ref{Halpha}. The separation between the emission peaks and the lack of a shell profile exclude an inclination angle close to 90 degrees, and hence the physical parameters in the first row of Table~\ref{fp} are not suitable. This implies that the rotational frequency is at least 90\% of the critical velocity. The shape of the H$\alpha$ line profile and the forbidden lines discussed above indicate the presence of a very extended circumstellar disk.

In spite of the high quality of the FEROS spectra and the presence of rapid photometric variability apparent in the CoRoT data (see Fig.~\ref{LC}), we have not been able to detect periodic variability in purely photospheric lines of HD 50\,209. Some variability is present slightly above the detection limit but no consistent periodic behavior could be established. Weak rapid variations were detected in the intensity of the $V$ and $R$ emission components of Balmer lines and their $V/R$ ratio. Note also that the quantity $(V+R)/2$ seems to be slightly lower in the second part of the ESO run, suggesting a slight weakening of the emissivity of the disk. As shown in the mean variance in the H$\beta$ profile, the variability is concentrated in the emission part. Note that the variability is more conspicuous on certain days. Unfortunately, the gap between the two parts of the ESO run, the alternating of day/night and the weakness of variations prevent any reliable estimate on the time-scale of variations.




\section{Discussion}\label{discussion}

As a result of our analysis, we have obtained 60 significant frequencies, grouped in six sets. In each set, different equidistant frequencies close to the main detected ones are present. The Fourier analysis of the light curve by means of sliding windows results in the detection of a significant variation of amplitude of the main frequencies (Fig.~\ref{3D}).

\begin{figure}
	\resizebox{\hsize}{!}{\includegraphics{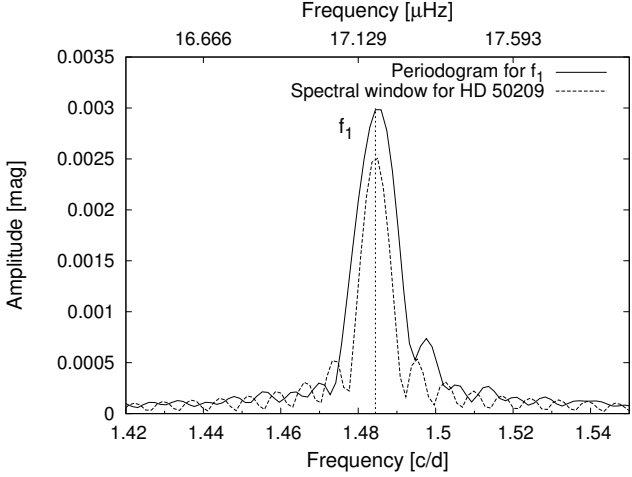}}
   	\caption{Periodogram of the original light curve (solid line) and spectral window shifted at $f_{1}=1.48444$ \cd (dashed line). Note that the full width at half maximum of the peak in the periodogram at the frequency $f_{1}$ is broader than expected from the spectral window, suggesting the presence of more frequencies or amplitude changes. The vertical line indicates the position of the frequency $f_{1}$.}
   	\label{FWHM1}
\end{figure}

The first issue to be addressed is the presence of frequency multiplets and amplitude variations. Both phenomena are related, and can be due either to the presence of close frequencies or actual amplitude variations \citep[][]{2006MNRAS.368..571B}. The beating of two or more close frequencies will appear as a single frequency with variable amplitude when we split the data sample in shorter intervals, with the consequent loss of frequency resolution. On the other hand, true amplitude variability will produce peaks in the power spectrum broader than expected from the frequency resolution (Fig.~\ref{FWHM1}), and the prewhitening of the main frequency will leave power in the wings of the main peak which will lead to false frequency doublets or multiplets.

In order to discriminate between the beating of several frequencies and an amplitude change of a single frequency we have applied the method described in \citet{2006MNRAS.368..571B} to study the relationship between amplitude variability and phase variations of an assumed single frequency. The idea behind this method is that the beating of frequencies will produce a phase variation with the same period as the beating period, while in the case of a single frequency with variable amplitude, the phase will remain constant.

We studied the six frequency groups with the abovementioned method. However, the results obtained are inconclusive, for two main reasons: i) the beat periods of the close frequencies are larger than the time coverage of CoRoT data, and hence, we cannot evaluate the consistency of the phase variations; ii) the phase variations predicted by the models with four or more frequencies are very small. At our detection level we were not able to discriminate between the predicted low amplitude variations and no variation at all. Consequently, we cannot firmly reject either of the two interpretations.

In the following, we analyze the six main frequencies found, which, as discussed above, can be either single frequencies with variable amplitude or groups of close frequencies around a central value. The highest frequency, $f_{6}$, is the first harmonic of $f_{1}$, i.e. $f_{6}=2\, f_{1}$, and hence it will not be considered in the analysis. Frequencies $f_{5}$, $f_{3}$, $f_{1}$ and $f_{2}$ are equidistant within the frequency resolution at the 3-$\sigma$ level, and the separation between them is the frequency $f_{4}$. Moreover, $f_{4}=0.67939$ \cd is consistent with the rotational frequency of the star as given in Table~\ref{fp}. As a consequence, we have only two independent main frequencies, $f_{4}$ and $f_{5}$, and all the others are of the form $f_{i}=f_{5}+n\, f_{4}$ with $n=1,2,3$.

Let us recall that the observed frequencies in a rotating star are related to the pulsational frequency in the co-rotating frame by the expression:
\begin{equation}
\nu=\,|\,\overline{\nu} - m \, \Omega\,| 
\end{equation}
where $\overline{\nu}$ is the frequency in the co-rotating frame and $\Omega$ is the rotational frequency of the star. From this expression, and considering $f_{4}$ as the rotational frequency as discussed above, the four frequencies $f_{5}$, $f_{3}$, $f_{1}$ and $f_{2}$ can be consistently interpreted as modes with the same pulsational frequency in the co-rotating frame and with $m = 0, -1, -2, -3$ respectively.

The fundamental parameters presented in Table~\ref{ground-based} place HD 50\,209 marginally outside the SPB instability strip calculated by \citet{1999AcA....49..119P}, although its position is compatible with the strip at the 1-$\sigma$ error level. Hence, we consider that the pulsations detected are gravity modes typical of SPB stars. HD 50\,209 is a SPBe star, the designation proposed by \citet{2005ApJ...635L..77W} for the Be stars pulsating in g-modes, considered as rapidly rotating counterparts of the SPB stars.

Due to the fact that in late-type B stars the frequencies of the g-modes in the co-rotating frame are much smaller than the rotational frequency, the frequencies of these modes in the observer's frame are close to $|\,m\, \Omega\,| $. This leads to the difficulty that the observed frequencies close to the expected rotational frequency can be either interpreted as g-mode pulsations \citep[][]{2005ApJ...635L..77W} or as rotational modulation \citep[][]{1995MNRAS.277.1547B}. In our case, we can for the first time discriminate between the rotational frequency and the frequency of the g-mode pulsation with azimuthal order $|\, m\, | =1$, as we have detected both of them clearly separated by an interval much larger than the frequency resolution, thanks to the long duration of observations and precise measurements of CoRoT. We can be confident that the frequencies $f_{5}$, $f_{3}$, $f_{1}$ and $f_{2}$ are not related to the rotation and hence they are true gravity mode pulsations.


The frequency in the co-rotating frame of the detected modes, namely $f_{5} = 0.10811$ \cd, is significantly lower than the frequencies commonly found for both SPB and SPBe stars. However, this low value is consistent with what is expected for an $m=0$ mode in models of rapidly rotating late-type B stars \citep[][]{2005ApJ...635L..77W,2007ApJ...654..544S}. The modelling of the detected pulsations will provide more insights into the nature of the pulsational modes.

\section{Conclusions}

The high precision photometry and long duration of continuous observations provided by the CoRoT mission has allowed the detection of g-mode pulsations in the late-type Be star HD 50\,209. This supports the fact that all Be stars have non-radial pulsations that could play a critical role in the mass ejection mechanism.

From our analysis, we have found pulsation in four modes with the same frequency in the co-rotating frame with azimuthal order $m=0,-1,-2,-3$. We have also detected the rotational frequency, both as a significant peak in the power spectrum and as the separation of frequencies with different $m$. The accurate determination of the rotational period will play an important part in  constraining the fundamental parameters of the star in order to perform the seismic modelling.


For the first time, we have been able to observe simultaneously the rotational frequency and the pulsational frequencies and separate them, implying that the frequencies we attribute to g-mode pulsations cannot be interpreted as the effect of the rotational modulation. This constitutes a proof of the presence of pulsations in HD 50\,209.



\begin{acknowledgements}
This research is based on CoRoT data. The CoRoT (Convection, Rotation and planetary Transits) space mission, launched on December 27th 2006, has been developed and is operated by CNES, with the contribution of Austria, Belgium, Brazil, ESA, Germany and Spain. We wish to thank the CoRoT team for the acquisition and reduction of the CoRoT data. The FEROS data have being obtained at ESO telescopes at the La Silla Observatory as part of the ESO Large Programme: LP178.D-0361 (PI: E. Poretti). The Narval data have been obtained at the T\'{e}lescope Bernard Lyot at Pic du Midi Observatory. This research has been financed by the Spanish ``Plan Nacional de Investigaci\'on Cient\'{\i}fica, Desarrollo e Innovaci\'on Tecnol\'ogica'', and FEDER, through contract AYA2007-62487. The work of P.~D. Diago is supported by a FPU grant from the Spanish ``Ministerio de Educaci\'{o}n y Ciencia''. The work of E.P., M.R. and K.U. was supported by the italian ESS project, contract ASI/INAF I/015/07/0, WP 03170. K.U. acknowledges financial support from a \textit{European Community Marie Curie Intra-European Fellowship}, contract number MEIF-CT-2006-024476.
\end{acknowledgements}


\bibliographystyle{aa}
\bibliography{11901.bib}

\Online

\onltab{2}{
\begin{table*}
\caption{Complete list of significant frequencies obtained with \textit{pasper} for the star HD 50\,209. The time reference for the phase is HJD = 2454391.95.}
\label{tablafreqs}
\centering
\begin{tabular}{c c c c c c c c c c}     
\hline\hline
\#	 & Freq.  		& Freq.	 	& Amp.		  & Amp. error	& Phase	 		& Phase error     & S/N   		   & Comment.\\
	 & [\cd]  		& [$\mu$Hz]	& [mmag]	  & [mmag]		& [0,1]	 		& [0,1]	 		  &	 			   &	   \\
\hline
1    &    1.48444   &  17.181   &   2.5299    &   0.0011    &   0.7348	    &	 0.0001   &    265      &  $f_{1}$     \\
2    &	  1.49028   &  17.2486  &   1.6167    &   0.0012    &   0.1577	    &	 0.0001   &    169      &       \\
3    &	  1.47860   &  17.1134  &   1.0901    &   0.0011    &   0.2747	    &	 0.0002   &    114      &       \\
4    &	  2.16238   &  25.0275  &   0.9043    &   0.0010    &   0.9746	    &	 0.0002   &    94       &  $f_{2}$     \\
5    &	  0.79482   &  9.19931  &   0.6077    &   0.0011    &   0.9866	    &	 0.0003   &    63       &  $f_{3}$     \\
6    &	  1.49613   &  17.3163  &   0.6531    &   0.0011    &   0.2873	    &	 0.0003   &    68       &       \\
7    &	  2.16822   &  25.0951  &   0.4953    &   0.0010    &   0.2636	    &	 0.0003   &    51       &       \\
8    &	  0.01461   &  0.16909  &   0.4324    &   0.0010    &   0.2149	    &	 0.0004   &    45       &  $<3/T$    \\
9    &	  0.77874   &  9.01319  &   0.4074    &   0.0010    &   0.0079	    &	 0.0004   &    42       &       \\
10   &	  0.80650   &  9.33449  &   0.3583    &   0.0010    &   0.4104	    &	 0.0005   &    37       &       \\
11   &	  0.67939   &  7.86331  &   0.4256    &   0.0010    &   0.3579	    &	 0.0004   &    44       &  $f_{4}$     \\
12   &	  0.69108   &  7.99861  &   0.3685    &   0.0009    &   0.5289	    &	 0.0004   &    38       &       \\
13   &	  0.82258   &  9.5206   &   0.2768    &   0.0010    &   0.4198	    &	 0.0006   &    29       &       \\
14   &	  0.80066   &  9.2669   &   0.2740    &   0.0011    &   0.1969	    &	 0.0006   &    28       &       \\
15   &	  0.10811   &  1.25127  &   0.2030    &   0.0010    &   0.5750	    &	 0.0008   &    21       &  $f_{5}$     \\
16   &	  0.81527   &  9.436    &   0.1725    &   0.0009    &   0.2895      &	 0.0009   &    18       &       \\
17   &	  0.78897   &  9.1316   &   0.1214    &   0.0011    &   0.8024	    &	 0.0014   &    12       &       \\
18   &	  0.09350   &  1.08218  &   0.1509    &   0.0009    &   0.4966	    &	 0.0010   &    15       &       \\
19   &	  0.03506   &  0.40578  &   0.1359    &   0.0009    &   0.3221	    &	 0.0011   &    14       &       \\
20   &	  0.11396   &  1.31898  &   0.1647    &   0.0010    &   0.7770	    &	 0.0010   &    17       &       \\
21   &	  1.51074   &  17.4854  &   0.1217    &   0.0009    &   0.2095	    &	 0.0012   &    12       &       \\
22   &	  0.65894   &  7.62662  &   0.1333    &   0.0010    &   0.2271	    &	 0.0011   &    13       &       \\
23   &	  2.96889   &  34.3622  &   0.1210    &   0.0009    &   0.7144	    &	 0.0012   &    12       &  $f_{6}$     \\
24   &	  0.07451   &  0.86238  &   0.0862    &   0.0009    &   0.2504	    &	 0.0017   &    9        &       \\
25   &	  0.64725   &  7.49132  &   0.0980    &   0.0009    &   0.1021	    &	 0.0015   &    10       &       \\
26   &	  0.82842   &  9.58819  &   0.1421    &   0.0010    &   0.7036	    &	 0.0011   &    14       &       \\
27   &	  0.00876   &  0.10138  &   0.1040    &   0.0010    &   0.6659	    &	 0.0015   &    10       &  $<3/T$     \\
28   &	  0.11980   &  1.38657  &   0.1164    &   0.0010    &   0.9496	    &	 0.0013   &    12       &       \\
29   &	  0.83573   &  9.6728   &   0.1160    &   0.0010    &   0.5225	    &	 0.0013   &    12       &       \\
30   &	  1.50197   &  17.3839  &   0.1158    &   0.0011    &   0.4101	    &	 0.0014   &    12       &       \\
31   &	  0.84157   &  9.74039  &   0.1131    &   0.0010    &   0.7923	    &	 0.0014   &    11       &       \\
32   &	  0.66916   &  7.74491  &   0.0847    &   0.0009    &   0.0830	    &	 0.0018   &    8        &       \\
33   &	  0.74952   &  8.675    &   0.0751    &   0.0009    &   0.0087	    &	 0.0019   &    7        &       \\
34   &	  0.04967   &  0.57488  &   0.1041    &   0.0009    &   0.5844	    &	 0.0014   &    10       &       \\
35   &	  0.70131   &  8.11701  &   0.0854    &   0.0009    &   0.8694	    &	 0.0017   &    8        &       \\
36   &	  0.02337   &  0.27048  &   0.0867    &   0.0009    &   0.2750	    &	 0.0017   &    9        &  $\sim3/T$	\\
37   &	  0.06136   &  0.71018  &   0.0894    &   0.0009    &   0.8152	    &	 0.0017   &    9        &       \\
38   &	  0.56397   &  6.52743  &   0.0792    &   0.0009    &   0.1145	    &	 0.0018   &    8        &       \\
39   &	  1.45668   &  16.8597  &   0.0774    &   0.0009    &   0.4253	    &	 0.0019   &    8        &       \\
40   &	  0.62533   &  7.23762  &   0.0694    &   0.0009    &   0.1556	    &	 0.0021   &    7        &       \\
41   &	  0.54205   &  6.27373  &   0.0726    &   0.0009    &   0.8328	    &	 0.0020   &    7        &       \\
42   &	  0.86203   &  9.9772   &   0.0731    &   0.0009    &   0.7081	    &	 0.0020   &    7        &       \\
43   &	  2.15361   &  24.926   &   0.0749    &   0.0009    &   0.2862	    &	 0.0019   &    7        &       \\
44   &	  0.72030   &  8.33681  &   0.0829    &   0.0009    &   0.8066	    &	 0.0018   &    8        &       \\
45   &	  2.96158   &  34.2775  &   0.0672    &   0.0009    &   0.4195	    &	 0.0021   &    7        &       \\
46   &	  0.73199   &  8.47211  &   0.0681    &   0.0009    &   0.0589	    &	 0.0022   &    7        &       \\
47   &	  0.12711   &  1.47118  &   0.0683    &   0.0009    &   0.1174	    &	 0.0021   &    7        &       \\
48   &	  2.97619   &  34.4466  &   0.0640    &   0.0009    &   0.0250	    &	 0.0022   &    6        &       \\
49   &	  2.17406   &  25.1627  &   0.0875    &   0.0010    &   0.5229	    &	 0.0019   &    9        &       \\
50   &	  2.17991   &  25.2304  &   0.0755    &   0.0010    &   0.7423	    &	 0.0021   &    7        &       \\
51   &	  1.53996   &  17.8236  &   0.0590    &   0.0009    &   0.4461	    &	 0.0024   &    6        &       \\
52   &	  0.77290   &  8.9456   &   0.0615    &   0.0010    &   0.8969	    &	 0.0025   &    6        &       \\
53   &	  1.38363   &  16.0142  &   0.0553    &   0.0009    &   0.7286	    &	 0.0026   &    5        &       \\
54   &	  0.47484   &  5.49583  &   0.0574    &   0.0009    &   0.4781	    &	 0.0025   &    6        &       \\
55   &	  0.61510   &  7.11921  &   0.0560    &   0.0009    &   0.2581	    &	 0.0026   &    5        &       \\
56   &	  1.52535   &  17.6545  &   0.0559    &   0.0009    &   0.2121	    &	 0.0026   &    5        &       \\
57   &	  0.87079   &  10.0786  &   0.0642    &   0.0009    &   0.8295	    &	 0.0023   &    6        &       \\
58   &	  0.88394   &  10.2308  &   0.0567    &   0.0009    &   0.9051	    &	 0.0026   &    5        &       \\
59   &	  0.58150   &  6.73032  &   0.0525    &   0.0009    &   0.6858	    &	 0.0028   &    5        &       \\
60   &	  0.17971   &  2.07998  &   0.0515    &   0.0009    &   0.1046	    &	 0.0028   &    5        &       \\
\hline
\end{tabular}
\end{table*}
}

\end{document}